\documentclass[12pt]{article}
\def\bV#1{\mbox{\boldmath $ #1 $}}

\begin{document}
\begin{center}
\begin{Large}
Gauge invariance of many-body Schr\"odinger equation \\ with explicit Coulomb potential \\[1cm]
\end{Large}
Kikuo Cho \\ 
Institute for Laser Engineering, Osaka University, Suita, Japan 
\end{center}

\begin{flushleft}
\underline{Abstract}  
\end{flushleft}
A simple argument is presented which, based on the minimal coupling Lagrangian for a many-body  
system, keeps the gauge invariance of the many-body Schr\"odinger equation with explicit Coulomb 
potential.  The elimination of longitudinal electric field does not necessarily lead to the 
breakdown of gauge invariance.  The total time derivative term in the matter-EM field interaction 
in the Lagrangian is canceled out by the choice of Coulomb gauge.  The remaining interaction 
is described by transverse vector potential and the longitudinal electric field which is the 
homogeneous solution of Gauss law.  This leads directly to the gauge invariant forms of the linear 
and nonlinear constitutive equations.  It is discussed how to reconcile this result with the cases 
of an isolated matter interacting with external charges.  \\

\section{Introduction}

Quantum mechanical description of electromagnetic (EM) response of matter 
requires the use of scalar and vector potentials for EM field.  
A general formulation starts usually 
from the so-called minimal coupling Lagrangian ${\cal L}$ for the system 
of charged particles and EM field.  It is the sum of particles Lagrangian 
and that of free EM field.  The particle part is a sum of single particle 
Lagrangian in a given EM field, $v^2/2m - e \phi + (e/c)\bV{v}\cdot\bV{A}$,  
over all the particles in consideration. (See eq.(\ref{eqn:L1}) for notations.) 
The minimum action principle leads to [a] Newton equation 
of motion of each particle under Lorentz force, and [b] Maxwell equations 
for EM field.  This fact guarantees the soundness of the Lagrangian.
A gauge transformation of the Lagrangian leads to an addition of a total time 
derivative to the original Lagrangian. Since the minimum action principle 
is not affected by such a term, we get a gauge invariant EM response.  

The Hamiltonian corresponding to ${\cal L}$ is derived via standard 
procedure, and the Schr\"odinger equation in terms of this Hamiltonian 
allows us to calculate the expectation values of various physical quantities. 
Among them, that of induced current density gives the constitutive equation, 
which, together with Maxwell equations, makes up the fundamental equations 
to calculate the EM response of matter.  In semiclassical theory where EM field 
is treated as classical (non-quantized) quantity, these fundamental equations 
are solved as simultaneous equations for a given initial condition of matter 
and EM field. Maxwell equations give EM field as a functional of current and 
charge densties, while constitutive equation gives current density as 
a functional of EM field.  

According to our standard conception about matter, the Coulomb potential 
among charged particles is a part of matter Hamiltonian.  The energy levels 
of a matter are given as the eigenvalues of the matter Hamiltonian consisting 
of the kinetic energies and the Coulomb potential of all the particles in the 
system, as typically seen in the Rydberg series of a hydrogen atom.  
However, ${\cal L}$ does not contain the Coulomb potential explicitly.  
It emerges from rewriting the self-energy of longitudinal (L) electric field 
in terms of Gauss law $\nabla\cdot\bV{E}= \rho/\epsilon_{0}$.  
When we derive the matter Hamiltonian with explicit Coulomb potential from 
the minimal coupling Lagrangian, the gauge invariance of the many-body 
Schr\"odinger equation, as mentioned in eqs.(\ref{eqn:G-tr}), (\ref{eqn:G-tr2}) 
below, is not automatically kept.  There seems to be no careful argument 
of this problem in literatures, to the author's knowledge.  Some people 
argue that gauge invariance is broken by eliminating the L electric field 
to derive Coulomb potential.  In this note, a simple argument is given 
to keep the gauge invariance in the many-body Schr\"odinger equation.

An additional point requiring a careful consideration is how we extract, 
from ${\cal L}$ for a global sysytem, the Lagrangian to describe the 
quantum mechanical motion of the isolated matter system (charged particles) 
of our interest.  For that purpose, we need to write ${\cal L}$  in terms 
of the internal and external variables, and after rewriting it into Hamiltonian 
with explicit Coulomb potential, we drop the irrelevant variables of the 
external origin.  Thereby, it is important to note 
that the L electric field in ${\cal L}$ contains, in adition to the chrage 
induced component, the solution $\bV{E}_{0}^{(\rm L)}$ of the homogeneous equation 
$\nabla\cdot\bV{E}=0$.  This term plays an essential role for the gauge 
invariance of the many-body Schr\"odinger equation.

Once we derive the matter Schr\"odinger equation in a gauge invariant form, the 
expectation value of an arbitrary physical quantity should be gauge-independent.  
Among all, Coulomb gauge is a special one in the sense that the Hamiltonian 
is already written in terms of the gauge independent components of EM field, 
i.e., the transverse (T) vector potential $\bV{A}^{(\rm T)}$ and homogeneous  
L electric field $\bV{E}_{0}^{(\rm L)}$. Hence the result 
obtained in Coulomb gauge directly gives the gauge independent form. 
This conclusion applies, not only to linear, but also to all the nonlinear 
EM responses.

\section{Formulation}

We start from the minimal coupling Lagarngian to describe a system of 
interacting charged particles and EM field as
\begin{eqnarray}
\label{eqn:L1}
  {\cal L} &=& \sum_{\ell}\ \{ \frac{1}{2} m_{\ell} v_{\ell}^2 
                         - e_{\ell} \phi(\bV{r}_{\ell})
            + e_{\ell} \bV{v}_{\ell}\cdot\bV{A}(\bV{r}_{\ell})\ \}  \nonumber \\
 &{}& \hspace{1cm} +  \frac{\epsilon_{0}}{2}\int {\rm d}\bV{r}\ \{\bV{E}^2 - c^2 \bV{B}^2 \}  \ ,
\end{eqnarray}
where $\bV{E} = -\partial \bV{A}/\partial t - \nabla \phi$, 
$\bV{B} = \nabla \times \bV{A}$, and 
$m_{\ell}, e_{\ell}, \bV{r}_{\ell}, \bV{v}_{\ell}$ are the mass, charge, 
coordinate, and velocity, respectively, of the $\ell$-th particle, and $\phi$ and 
$\bV{A}$ scalar and vector potential, respectively. 
The interaction terms in ${\cal L}$ can be rewritten as
\begin{equation}
   \sum_{\ell}\ \{- e_{\ell} \phi(\bV{r}_{\ell}) + 
   e_{\ell} \bV{v}_{\ell}\cdot\bV{A}(\bV{r}_{\ell})\ \} 
   = \int{\rm d}\bV{r}\ \{-\rho(\bV{r}) \phi(\bV{r}) 
           + \bV{J}(\bV{r}) \cdot \bV{A}(\bV{r})\} \ ,
\end{equation}
where  
\begin{eqnarray}
  \rho(\bV{r}) &=& \sum_{\ell} e_{\ell}\ \delta(\bV{r} - \bV{r}_{\ell}) \ , \\
  \bV{J}(\bV{r}) &=& \sum_{\ell} e_{\ell}\ \bV{v}_{\ell}\ \delta(\bV{r} - \bV{r}_{\ell}) 
\end{eqnarray}
are charge and current densities, respectively. 

As well known, the Lagrange equations due to this ${\cal L}$ lead to the 
Newton equation of motion of the $\ell$-th particle 
\begin{equation}
  m_{\ell} \frac{d^2\bV{r}_{\ell}}{dt^2} =  e_{\ell} \{\bV{E}(\bV{r}_{\ell}) 
                      + \bV{v}_{\ell} \times \bV{B}(\bV{r}_{\ell}) \} \ ,
\end{equation}
and the Maxwell equations for EM field
\begin{eqnarray} 
\label{eqn:Gauss}
  &{}& \nabla \cdot [\frac{\partial \bV{A}}{\partial t} + \nabla \phi] = -\frac{\rho}{\epsilon_{0}}
          \ \ \ \ (\nabla \cdot \bV{E} = \frac{\rho}{\epsilon_{0}})  \ , \\
\label{eqn:wave}
  &{}& \frac{1}{c^2} \frac{\partial ^2 \bV{A}}{\partial t^2} - \nabla^2 \bV{A} 
         + \nabla \big( \nabla\cdot \bV{A} + \frac{1}{c^2} \frac{\partial \phi}{\partial t} \big)
         = \mu_{0} \bV{J} \ .         
\end{eqnarray}
Maxwell equations give EM field as a functional of $\rho\ {\rm and}\ \bV{J}$, which also contains 
the solution of the homogeneous equations (for $\rho = 0, \ \bV{J}=0$).  The existence of 
the field component independent of charge density is important for the gauge invariant 
rewriting of ${\cal L}$ into many-body Hamitonian with explicit Coulomb potential. 


Now we rewrite ${\cal L}$ by using Gauss law, $\nabla \cdot \bV{E}^{(\rm L)} 
=  \rho / \epsilon_{0}$, which leads to Coulomb potential and eliminates the 
time derivative of $\bV{A}^{(\rm L)}$ from ${\cal L}$.  The general solution of 
Gauss law is the sum of the charge induced field ${\cal E}$ and the solution 
$\bV{E}_{0}^{(\rm L)}$ of the homogeneous equation as 
\begin{eqnarray}
\label{eqn:EL}
  \bV{E}^{(\rm L)}(\bV{r}) &=& \bV{E}_{0}^{(\rm L)} + {\cal E}(\bV{r}) \ , \\ 
\label{eqn:CoulF}
{\cal E}(\bV{r}) &=& 
 - \frac{1}{4\pi \epsilon_{0}} \nabla \int{\rm d}\bV{r}'\ \frac{\rho(\bV{r}')}{|\bV{r}-\bV{r}'|} 
\end{eqnarray}
where $\bV{E}_{0}^{(\rm L)}$ is a spatially constant vector satisfying 
$\nabla \cdot \bV{E}_{0}^{(\rm L)} = 0$.  Its representation in terms of scalar 
and vector potentials 
$\bV{E}^{(\rm L)}_{0} = -(\partial \bV{A}^{(\rm L)}_{0}/\partial t) - \nabla \phi_{0} $
has infinite choices according to gauge transformation. 

Since we are interested in an isolated matter system, we divide $\rho$ into 
the internal and external components 
\begin{equation}
\label{eqn:rho-ie}
  \rho = \rho_{\rm int} + \rho_{\rm ext}
\end{equation}
where $\rho_{\rm int}$ belongs to the isolated system.  Correspondingly, 
the charge induced field is also the sum of the components induced by 
$\rho_{\rm int}$ and $\rho_{\rm ext}$
\begin{equation}
  {\cal E} ={\cal E}_{\rm int} + {\cal E}_{\rm ext} \ .
\end{equation}

The $\bV{E}^2$ term of ${\cal L}$ can be divided into T and L components as  
\begin{equation}
  \frac{\epsilon_{0}}{2} \int{\rm d}\bV{r} [\bV{E}^{(\rm L)}(\bV{r})^2 + \bV{E}^{(\rm T)}(\bV{r})^2] \ .
\end{equation}
The integral of the L component is evaluated as   
\begin{eqnarray}
   \frac{\epsilon_{0}}{2}\int {\rm d}\bV{r}\ [\bV{E}^{(\rm L)}]^2 &=& 
       \frac{\epsilon_{0}}{2}\int {\rm d}\bV{r}\ 
  [\{\bV{E}_{0}^{(\rm L)}\}^2 + {\cal E}_{\rm int}^2 + {\cal E}_{\rm ext}^2 \nonumber \\
   &+& 2 {\cal E}_{\rm int} \cdot {\cal E}_{\rm ext} 
    + 2 \bV{E}_{0}^{(\rm L)} \cdot \{{\cal E}_{\rm int} + {\cal E}_{\rm ext}\} ] \ .
\end{eqnarray}
The terms 
\begin{equation}
\frac{\epsilon_{0}}{2}\int {\rm d}\bV{r}\ 
  [{\cal E}_{\rm int}^2 + 2 {\cal E}_{\rm int} \cdot {\cal E}_{\rm ext}
    + 2 \bV{E}_{0}^{(\rm L)} \cdot {\cal E}_{\rm int} ]    
\end{equation}
among others influence the motion of internal particles via variational principle.
Using (\ref{eqn:CoulF}), we can evaluate the integral as a sum of the Coulomb 
potentials 
\begin{eqnarray}
  U_{\rm C} &=& \frac{1}{8\pi\epsilon_{0}} \int\int{\rm d}\bV{r}{\rm d}\bV{r}' \ 
     \frac{\rho_{\rm int}(\bV{r}) \rho_{\rm int}(\bV{r}')}{|\bV{r} - \bV{r}'|} \ , \\
  U_{\rm ie} &=& \frac{1}{4\pi\epsilon_{0}} \int\int{\rm d}\bV{r}{\rm d}\bV{r}'\  
     \frac{\rho_{\rm int}(\bV{r}) \rho_{\rm ext}(\bV{r}')}{|\bV{r} - \bV{r}'|}    
\end{eqnarray}
The integral containing $\bV{E}_{0}^{(\rm L)}$ is written as 
\begin{equation}
   - \frac{1}{4\pi} \int\int{\rm d}\bV{r}{\rm d}\bV{r}' \ \bV{E}_{0}^{(\rm L)} 
        \cdot \nabla  \frac{\rho_{\rm int}(\bV{r}')}{|\bV{r} - \bV{r}'|} \ ,
\end{equation}
which is zero by using $\nabla\cdot\bV{E}_{0}^{(\rm L)} = 0$ and the fact that 
the upper and lower limits of the integral can be taken arbitrarily large outside 
the isolated matter system. 

Eliminating the variables irrelevant to the motion of internal particles, we 
obtain the Lagrangian for the internal particles 
\begin{eqnarray}
\label{eqn:Lin}
  {\cal L}_{\rm i} &=& \sum_{\ell}\ \frac{1}{2} m_{\ell} v_{\ell}^2 
                         + \int{\rm d}\bV{r}\ \{-\rho_{\rm int}(\bV{r}) \phi(\bV{r}) 
           + \bV{J}_{\rm int}(\bV{r}) \cdot \bV{A}(\bV{r})\} \nonumber \\
   &{}& + U_{\rm C} + U_{\rm ie} 
         +  \frac{\epsilon_{0}}{2}\int {\rm d}\bV{r}\ 
         \{[\frac{\partial \bV{A}^{(\rm T)}}{\partial t}]^2 - [\nabla \times \bV{A}^{(\rm T)}]^2\}\ .
\end{eqnarray}
All the particle variables, except for those of $U_{\rm ie}$, are the ones for 
internal system.  Due to the rewriting to make Coulomb potentials explicit, 
$\bV{A}^{(\rm L)}$ appears only in the interaction term.  The signs of $U_{\rm C}$ and 
 $U_{\rm ie}$ are positive here, but there arise cancelling contributions from the 
interaction terms.

Using $\bV{J}_{\rm int}^{(\rm L)} = \partial\bV{P}_{\rm int}^{(\rm L)} / \partial t$ , 
we can rewrite the L-field related interaction terms as 
\begin{eqnarray}
&{}&  \int{\rm d}\bV{r}[ -  \rho_{\rm int}(\bV{r}) \phi(\bV{r}) 
                    +  \bV{J}_{\rm int}^{(\rm L)}(\bV{r})\cdot \bV{A}^{(\rm L)}(\bV{r})] \nonumber \\
&{}& \hspace{2cm}  =   \int{\rm d}\bV{r}  \bV{P}_{\rm int}^{(\rm L)} \cdot \bV{E}^{(\rm L)}  
  +  \frac{d}{dt} \int{\rm d}\bV{r} \bV{P}_{\rm int}^{(\rm L)} \cdot \bV{A}^{(\rm L)} \ ,
\end{eqnarray}
where $\nabla \cdot \bV{P}_{\rm int}^{(\rm L)} = - \rho_{\rm int}$.  
The first term of the r.h.s. can be further written, by the help of  
eqs.(\ref{eqn:EL} - \ref{eqn:rho-ie}), as 
\begin{equation}
 \int{\rm d}\bV{r}  \bV{P}_{\rm int}^{(\rm L)} \cdot \bV{E}^{(\rm L)} = -2U_{\rm C} - U_{\rm ie} 
  +   \int{\rm d}\bV{r}  \bV{P}_{\rm int}^{(\rm L)} \cdot \bV{E}_{0}^{(\rm L)}  .
\end{equation}
This allows us to rewrite ${\cal L}_{\rm i}$ as 
\begin{eqnarray}
\label{eqn:Lagrange3}
  {\cal L}_{\rm i} &=& \sum_{\ell}\ \frac{1}{2} m_{\ell} v_{\ell}^2 
        + \int{\rm d}\bV{r}\ \bV{J}_{\rm int}^{(\rm T)}(\bV{r}) \cdot \bV{A}^{(\rm T)}(\bV{r}) 
     + \int{\rm d}\bV{r}  \bV{P}_{\rm int}^{(\rm L)} \cdot \bV{E}_{0}^{(\rm L)} - U_{\rm C} \nonumber \\
   &{}& +  \frac{d}{dt} \int{\rm d}\bV{r} \bV{P}_{\rm int}^{(\rm L)} \cdot \bV{A}^{(\rm L)} 
         +  \frac{\epsilon_{0}}{2}\int {\rm d}\bV{r}\ 
         \{[\frac{\partial \bV{A}^{(\rm T)}}{\partial t}]^2 - [\nabla \times \bV{A}^{(\rm T)}]^2\}\ .
\end{eqnarray}
The minimum action principle is not affected by the total time derivative term in Lagrangian. 
The possibility of its elimination means that the EM field variables do not contain the L field 
induced by $\rho_{\rm ext}$. 

Before showing how to eliminate the total time derivatve term, we point out the gauge invariance 
of the many particle Schr\"odinger equation based on eq.(\ref{eqn:Lin}). 
The generalized momenta are 
\begin{equation}
 \bV{p}_{\ell} = m_{\ell} \bV{v}_{\ell} + e_{\ell} \bV{A}(\bV{r}_{\ell}) \ , \ \ \
 \bV{\Pi}_{\rm AT} = \epsilon_{0} \frac{\partial \bV{A}^{(\rm T)}}{\partial t} \ ,
\end{equation}
and the corresponding Hamiltonian is 
\begin{eqnarray}
  {\cal H} &=& \sum_{\ell} \bV{p}_{\ell} \cdot \bV{v}_{\ell} 
   + \int{\rm d}\bV{r} \ \bV{\Pi}_{\rm AT}\cdot \frac{\partial \bV{A}^{(\rm T)}}{\partial t} 
   - {\cal L}_{\rm i}   \nonumber \\
   &=& \sum_{\ell}\frac{1}{2m_{\ell}}[\bV{p}_{\ell} - e_{\ell}\bV{A}(\bV{r}_{\ell})]^2 
       - U_{\rm C} - U_{\rm ie} + \int{\rm d}\bV{r}\ \rho_{\rm int}(\bV{r}) \phi(\bV{r}) \nonumber \\
   &{}& + \frac{\epsilon_{0}}{2}\int {\rm d}\bV{r}\ 
         \{[\frac{\partial \bV{A}^{(\rm T)}}{\partial t}]^2 + [\nabla \times \bV{A}^{(\rm T)}]^2\}\ .
\end{eqnarray} 
The last term is the Hamiltonian of transverse EM field (vacuum field) and the rest is 
the matter Hamiltonian in a given EM field 
\begin{equation}
\label{eqn:Hamilton4}
  {\cal H}_{\rm M} = \sum_{\ell}\  \frac{1}{2 m_{\ell}} 
                      \{ \bV{p}_{\ell} - e_{\ell} \bV{A}(\bV{r}_{\ell}) \}^2 
                 - U_{\rm C} - U_{\rm ie} +  \int{\rm d}\bV{r}  \rho(\bV{r}) \phi(\bV{r}) \ .
\end{equation}
The Schr\"odinger equation of the internal particles in a given EM field is   
\begin{equation}
\label{eqn:Seq}
  i\hbar \frac{\partial \Psi}{\partial t} = {\cal H}_{\rm M} \Psi 
\end{equation}
and its solution allows us to calculate the expectation value of any physical quantities, 
especially the induced current density, which plays the role of the constitutive equation 
to determine the EM response together with Maxwell equations.  

It should be noted that this many-body Schr\"odinger equation is gauge invariant, i.e., 
the gauge transformation  $\{\bV{A}, \phi, \Psi \} \rightarrow \{\bV{A}', \phi', \Psi' \}$ 
mediated by an arbitrary scalar function $\chi(\bV{r}, t)$  
\begin{eqnarray}
\label{eqn:G-tr}
 \bV{A}'&=& \bV{A} + \nabla \chi \ , \ \ \ 
 \phi' = \phi - \frac{\partial \chi}{\partial t}\ ,  \ \ \ \Psi' = \exp(i\Theta) \Psi \ , \\
\label{eqn:G-tr2}
   \Theta &=& \sum_{\ell} \frac{e_{\ell}}{\hbar}\chi(\bV{r}_{\ell}, t)  \ 
             = \frac{1}{\hbar} \int{\rm d}\bV{r}\ \rho_{\rm int}(\bV{r}) \chi(\bV{r},t) \ .
\end{eqnarray}
does not change the Schr\"odinger equation.  If we denote the coordinate, velocity, 
and spin of an arbitrary particle as $\hat{\cal O}$, we have  
$\hat{\cal O}  \Psi' = \exp(i\Theta) \hat{\cal O} \Psi$, so that the expectation value 
\begin{equation}
  \langle\Psi'| f(\hat{\cal O}) |\Psi' \rangle =  \langle\Psi| f(\hat{\cal O}) |\Psi \rangle
\end{equation}
of any physical quantity written as a function of $\hat{\cal O}$'s is gauge invariant.  

When we treat materials containing heavy elements, or magnetic species, etc., it is 
required to consider the relativistic corrections, such as spin-orbit interaction 
(${\cal H}_{\rm so}$), spin Zeeman interaction (${\cal H}_{\rm Z}$), etc.  For that 
purpose, we simply add the corresponding terms to ${\cal H}_{\rm C}$.  The spin Zeeman 
term can be included in the matter-EM field interaction term, and the rest in the matter 
Hamiltonian ${\cal H}_{\rm M}$. This correction does not change the argument about the 
gauge invariance, since the correction terms are written in terms of $\bV{E}$ and $\bV{B}$ 
\cite{Schiff, Frohlich}. \\

\section{Discussions}
\subsection{Special meaning of Coulomb gauge}

In order to eliminate the total time derivative term in the ${\cal L}_{\rm i}$, we simply  
need to make a gauge transformation to cancell it, which turns out to be Coulomb gauge, as 
shown below. If we apply the gauge transformation (\ref{eqn:G-tr}) to ${\cal L}_{\rm i}$, 
the difference $\delta{\cal L}_{\rm i}$ arises only from the interaction terms as 
\begin{equation}
\label{eqn:deltaL}
  \delta {\cal L}_{\rm i} = \ \int{\rm d}\bV{r} 
                      [\bV{P}_{\rm int}^{(\rm L)} \cdot \nabla \frac{\partial \chi}{\partial t} 
                            + \frac{\partial \bV{P}_{\rm int}^{(\rm L)}}{\partial t} \cdot \nabla \chi ] 
                       =  \frac{d}{dt}  \int{\rm d}\bV{r}\ \bV{P}_{\rm int}^{(\rm L)} \cdot \nabla \chi  \ .    
\end{equation}
If we choose $\chi$ satisfying $\nabla \chi = - \bV{A}^{(\rm L)}$, the total time derivative term 
of (\ref{eqn:Lagrange3}) is canceled.  The choice $\nabla \chi = - \bV{A}^{(\rm L)}$ eliminates 
the L component of $\bV{A}$, so that it is equivalent to the choice of Coulomb gauge. 

Thus, the Lagrangian in Coulomb gauge is 
\begin{eqnarray}
\label{eqn:Lagarnge4}
  {\cal L}_{\rm C} &=& \sum_{\ell}\ \frac{1}{2} m_{\ell} v_{\ell}^2  - \ U_{\rm C} 
                 +  \int{\rm d}\bV{r}  \bV{P}^{(\rm L)} \cdot \bV{E}_{0}^{(\rm L)} 
                 + \int{\rm d}\bV{r} \bV{J}^{(\rm T)}\cdot \bV{A}^{(\rm T)}       \nonumber \\
 &{}&          + \frac{\epsilon_{0}}{2}\int {\rm d}\bV{r}\ \left[ 
                  (\frac{\partial \bV{A}^{(\rm T)}}{\partial t} )^2 
                                - c^2 (\nabla \times \bV{A}^{(\rm T)})^2 \right] \ . 
\end{eqnarray}
and the corresponding Hamiltonian is 
\begin{eqnarray}
\label{eqn:Hamilton2}
  {\cal H}_{\rm C} &=& \sum_{\ell}\  \frac{1}{2 m_{\ell}} 
             \{ \bV{p}_{\ell} - e_{\ell} \bV{A}^{(\rm T)}(\bV{r}_{\ell}) \}^2 
     + U_{\rm C} - \int{\rm d}\bV{r}  \bV{P}^{(\rm L)} \cdot \bV{E}_{0}^{(\rm L)}  \nonumber \\
     &{}& \hspace{2cm}         +  \frac{\epsilon_{0}}{2}\int {\rm d}\bV{r}\ \{ 
            (\frac{\partial \bV{A}^{(\rm T)}}{\partial t})^2 
                                      + c^2 (\nabla \times \bV{A}^{(\rm T)})^2 \}  \ .
\end{eqnarray}

It should be stressed that the EM variables in this Hamiltonian are  $\bV{A}^{(\rm T)}$ 
and $\bV{E}_{0}^{(\rm L)}$, which are the gauge independent components of EM field. 
Combining the facts that ${\cal H}_{\rm i}$ gives the gauge invariant Schr\"odinger equation 
and that ${\cal H}_{\rm C}$ is written only in terms of gauge independent components of 
EM field, we directly get the gauge invariant EM response from the Schr\"odinger equation 
in terms of ${\cal H}_{\rm C}$. 
The induced current density is given as a power series expansion with respect to these 
field variables, and their coefficients, i.e., the susceptibilities, are written in terms 
of the eigenvalues and eigenfunctions of the many-body Hamiltonian 
${\cal H}_{\rm mb} = \sum_{\ell} (p_{\ell}^2/2m_{\ell}) + U_{\rm C}$ (including, if 
necessary, relativistic corrections).  The merit of this 
representation is that the susceptibilities are given as separable integral kernels, 
which plays an essential role in solving integral equations and also carrying out the long 
wavelength approximation to derive macroscopic constitutive equation \cite{Cho1, Cho2}.  
The arguments given above enforces the foundation of these micro- and macroscopic 
response theories by assuring the gauge invariant nature.

\subsection{Homogeneous L field vs. "External" L field} 
 
The L electric feld $\bV{E}_{0}^{(\rm L)}$, the solution of homogeneous Gauss law, 
has turned out to play an essential role in the gauge invariance of many-body 
Schr\"odinger equation.  However, the fact that it has nothing to do with internal 
and external charge densities makes us wonder how it is, or is not, related with the 
description of a matter sample disturbed by an external charge density in cases, 
such as the polarization of a matter sample placed in a condencer, or energy loss 
spectroscopy of an electron beam incident on a matter sample, etc.  In these examples, 
the electric field due to charged condenser or electron beam is treated as an 
external electric field inducing a L polarization in the sample.  This type of 
analyses are known to work well, but the problem here is how it fits to the 
gauge invariant formalism. 

Obviously, we cannot ascribe the L electric field arising from charged condenser or 
electron beam to $\bV{E}_{0}^{(\rm L)}$.  We should rather leave $\bV{E}_{0}^{(\rm L)}$ 
as a free L field and regard both of the matter sample and the charges inducing 
the L field as the "internal system" of the previous sections.  In this way we can 
keep the gauge invariant many-body formulation.  The gauge to describe this internal 
system is arbitrary, but Coulomb gauge will be of pactical convenience in many cases.  
Dividing the internal system into two parts, [A] sample and [B] charged particles 
on the condenser or electron beam, we consider the coupling of the two parts for 
a given initial condition.  In this scheme, we should generally consider the mutual 
action and reaction between [A] and [B]. 
In the case of condenser, it polarizes the matter sample and its polarization changes 
the capacity of the condenser affecting the charges accumulated on the condenser. 
In the case of electron energy loss spectroscopy, electron beam induces the matter 
excitation with L polarization, plasmons for example, and the corresponding change 
occurs in the electron beam, showing the energy and momentum loss corresponding 
matter excitation. In the latter case, the energy loss function for the beam can be 
represented by the inverse of L dielectric function $1/\epsilon(\bV{k}, \omega)$ of 
the matter \cite{Pines}.

\subsection{Problem of "velocity vs. length gauge"}

In the quantum mechanical description of matter - EM field interaction, there have been 
a lot of arguments as to the form of interaction between charged particle and EM field,  
either $-\bV{E}\cdot \bV{r}$ or $(-e/m)\bV{p}\cdot\bV{A}$, since early days \cite{velolength}.  
They are said to correspond to different gauges of EM field, and are usually called 
length gauge and velocity gauge, respectively.  

These two forms of interaction are connected by Power-Zienau-Woolley (PZW) transformation 
in the Lagrangian \cite{Cohen}, which is specified by the addition, to the 
Lagrangian in the Coulomb gauge, of the total time derivative of 
\begin{equation}
 F = -\int{\rm d}\bV{r} \bV{P}(\bV{r}) \cdot \bV{A}(\bV{r}) \ .
\end{equation}
The new Lagrangian equally serves to the quantum mechanical description of the system. 
Through this addition, the interaction term 
\begin{equation}
  {\cal L}_{\rm int} = \int{\rm d}\bV{r} \bV{J}(\bV{r}) \cdot \bV{A}(\bV{r}) \ ,
\end{equation}
becomes 
\begin{equation}
  {\cal L}'_{\rm int} = \int{\rm d}\bV{r}\{ 
                   \bV{P}^{(\rm T)}(\bV{r}) \cdot \bV{E}^{(\rm T)}(\bV{r}) 
                 + \bV{M}(\bV{r}) \cdot \bV{B}(\bV{r}) \}
\end{equation}
via the identity $\bV{J} = (\partial \bV{P}/\partial t) + \nabla \times \bV{M}$. 
The interaction term $\bV{J} \cdot \bV{A}$ stands for the "velocity gauge" 
and $\bV{P}^{(\rm T)}\cdot \bV{E}^{(\rm T)}$ for the "length gauge". 

It is instructive to compare the total time derivative term $dF/dt$ with the similar  
term caused by a gauge transformation, eq.(\ref{eqn:deltaL}).  For any choice of 
$\chi$ they cannot be same, because the T component $\bV{P}^{(\rm T)}$ exists in 
$F$ but not in the latter.  This shows that the nonuniqueness of a Lagrangian with 
respect to the addition of an arbitrary total time derivative term is a broader 
concept than the one due to gauge transformation.  (This is reasonable from a more 
general viewpoint, i.e.,  Lagrangian formalism works also in systems which have 
nothing to do with electromagnetism.)  For this reason, "$-\bV{E}\cdot \bV{r}$ or  
$(-e/m)\bV{p}\cdot\bV{A}$" is not a problem of gauge.

\section{Summary}
In summary, we have shown that the many-body Schr\"odinger equation with explicit 
Coulomb potential can be given in a gauge invariant form, and that the EM response 
obtained in Coulomb gauge directly gives the gauge invariant one. \\

This work is supported in part by Grant-in-Aid for scientific research (Grant-No. 
No.25610071) and Innovative Areas Electromagnetic Metamaterials (Grant No. 22109001) 
of MEXT, Japan.


\begin{thebibliography}{9}
\bibitem {Schiff} L. I. Schiff: {\it Quantum Mechanics}, McGraw-Hill New York 1955, 
                  Chap.XII 
\bibitem {Frohlich} J. Fr\"ohlich and U. M. Studer: Rev. Mod. Phys. {\bf 65} (1993) 733                 
\bibitem {Cho1} K. Cho, {\it Optical Response of Nanostructures: Microscopic 
                                 Nonlocal Theory}, Springer Verlag 2003
\bibitem {Cho2} K. Cho, {\it Reconstruction of Macroscopic Maxwell Equations: 
                   A Single Susceptibility Theory}, Springer Verlag 2010 ; 
                   J. Phys.: Condens. Matter {\bf 20} (2008) 175202  
\bibitem {Pines} D. Pines, {\it Elementary excitations in solids}, Sec.3.4, Benjamin 
                 New York, 1964  
\bibitem {velolength} W.E.Lamb, Jr., R. R. Schlichter, and M. O. Scally, Phys. Rev. 
                  A{\bf 36} (1987);  R. Del Sole and R. Girtlanda, Phys. Rev. B{\bf 48} 
                 (1993) 11789; Scully and Zubairy, {\it Quantum Optics} Appendix 5A, 
                  Cambridge University Press 1997    
\bibitem {Cohen} C. Cohen-Tannoudji, J. Dupont-Roc, and G. Grynberg, 
                 {\it Photons and Atoms}, Sec.IV.C, (Wiley Interscience, New York 1989)                 
\end{thebibliography}
\end{document}